\documentclass[%
 reprint,
 amsmath,amssymb,
 aps,
prb,
]{revtex4-2}

\usepackage{graphicx}
\usepackage{dcolumn}
\usepackage{bm}
\usepackage{braket}
\usepackage{changes}
 \usepackage{color}
\usepackage{comment}
\usepackage{hyperref}       
\hypersetup{
    colorlinks=true,
    linkcolor=blue,
    filecolor=magenta,      
    urlcolor=blue,
}
\usepackage{color,soul}

\begin{document}

\preprint{APS/1}

\title{On-chip topological transport of optical frequency combs in silicon-based valley photonic crystals}

\author{Zhen Jiang$^{1,2,\#}$}%
\author{Hongwei Wang$^{1,\#}$}%
\author{Yuechen Yang$^{1,2}$}%
\author{Yang Shen$^{1,2}$}%
\author{Bo Ji$^{1,2}$}%
\author{Yanghe Chen$^{1,2}$}%
\author{Yong Zhang,$^{1}$}%
\author{Lu Sun,$^{1}$}%
\author{Zheng Wang,$^{1}$}%
\author{Chun Jiang$^{1}$}%
\email{cjiang@sjtu.edu.cn}	
\author{Yikai Su$^{1}$}%
\email{yikaisu@sjtu.edu.cn}	
\author{Guangqiang He$^{1,2}$}%
\email{gqhe@sjtu.edu.cn}	
\affiliation{%
 $^1$State Key Laboratory of Advanced Optical Communication Systems and Networks, Department of Electronic Engineering, Shanghai Jiao Tong University, Shanghai 200240, China\\
 $^2$SJTU Pinghu Institute of Intelligent Optoelectronics, Department of Electronic Engineering, Shanghai Jiao Tong University, Shanghai 200240, China \\
 $^\#$These authors contributed equally to this work.
}%
	
\date{\today}


\maketitle

\textbf{The generation and control of optical frequency combs in integrated photonic systems enables complex,  high-controllable, and large-scale devices.
In parallel, harnessing topological physics in multipartite systems has allowed them with compelling features such as robustness against fabrication imperfections.
Here we experimentally demonstrate on-chip topological transport for optical frequency combs at telecommunication wavelengths, both in classical and nonclassical domains.  
We access both the quantum frequency combs and dissipative Kerr soliton combs with a micro-resonator. The quantum frequency comb, that is, a coherent superposition of multiple frequency modes, is proven to be a frequency-entangled qudit state.
We also show that dissipative Kerr soliton combs are highly coherent and mode-locked due to the collective coherence or self-organization of solitons. 
Moreover, the valley kink states allow both quantum frequency combs and dissipative Kerr soliton combs with robustness against sharp bends. Our topologically protected optical frequency combs could enable the inherent robustness in integrated complex photonic systems.}

\section{Introduction}

Integrated optical frequency combs (OFCs) in classical and quantum domains are undergoing a revolutionary development due to the advances in nanofabrication techniques. 
OFCs characterized by phase-coherent, uniformly spaced narrow laser lines can greatly extend the frequency domain channels, therefore, they have a great potential to increase the number of effective dimensions of photonic systems~\cite{1-1,1-2,1-3}. 
As a type of OFCs, quantum frequency combs (QFCs) generated at the single-photon level are particularly important for quantum information science due to their high-dimensional distributions in the frequency and time domains~\cite{1-4,1-5,1-6}. QFCs have demonstrated high-dimensional frequency entanglement~\cite{1-7,1-8,1-9}, energy-time entanglement~\cite{1-10,1-11,1-12}, and time-bin multiphoton entanglement~\cite{1-13}, which provide practical applications in quantum communication~\cite{1-14} and quantum computing~\cite{1-15}.
For classical frequency combs, mode-locking approaches provide high stability for fully coherent dissipative Kerr soliton (DKS) combs, which have been used in ultrafast ranging~\cite{2-1,2-2,2-3,2-4}, optical communications~\cite{2-5}, optical spectroscopy~\cite{2-6,2-7}, frequency synthesis~\cite{2-8,2-9}, and optical computing~\cite{2-10,2-11}. However, the self-coherent frequency combs, especially QFCs, are fragile due to decoherence and fidelity degradation in noisy environments~\cite{3-1,3-2}.

Meanwhile, topological photonics has paved the way for robust optical transmission utilizing topological protection, which can be categorized into quantum Hall effect~\cite{3-2-1,3-2-2,3-2-3}, quantum spin Hall effect (QSH)~\cite{3-3,3-4,3-5,3-6} and quantum valley Hall effect (QVH)~\cite{3-8,3-9,3-10,3-11}. 
They have given rise to new paradigms of optical functional devices characterized by directional transmission and high interference immunity. 
Recently, significant advances have been made in topological nonlinear and quantum photonics, including topological nonlinear harmonic generation~\cite{3-12,3-13}, topological nonlinear imaging~\cite{3-14}, topological single-photon and biphoton states~\cite{3-15,3-16,3-17}, topological quantum emitters~\cite{3-2}, and topological quantum interference~\cite{3-18,3-19}. 
Topological quantum entanglement in coupled resonator arrays emulating the Su–Schrieffer–Heeger (SSH) model has been experimentally proposed~\cite{3-2}. However, the coupled resonator arrays have large chip sizes and are inconvenient for optical modulation. 
The topological transport in multiparty and high-dimensional entangled QFCs remains to be established. The valley photonic crystal (VPC) waveguides have advantages in large bandwidth and super-compact chip sizes. Hence they are convenient for optical modulation, offering exciting possibilities for on-chip robust topological transport of high-dimensional entangled QFCs.

Here we experimentally demonstrate on-chip topological transport for both QFCs and DKS combs generated from the same silicon nitride ($\rm Si_{3}N_{4}$) micro-resonator. A VPC-based waveguide supporting topologically protected kink states is fabricated on a silicon-on-insulator (SOI) slab. 
Employing a $\rm Si_{3}N_{4}$ micro-resonator with an approximate 100 GHz free spectral range (FSR), we can access QFCs and DKS combs by pumping different powers. On the one hand, we measure the coincidence-to-accidental ratio (CAR) and joint spectral intensity (JSI) of QFCs after their transport to topological interfaces. By calculating the Schmidt numbers of the QFC, we show that it remains a two-photon entangled qudit state. This indicates that the entangled biphootn state is robust against structural imperfections. On the other hand, we also access DKS combs operating in single soliton, multiple solitons, and soliton crystals, respectively. Although these DKS combs traverse the topological interfaces, they still retain their distinctive DKS-like characteristics. Topologically protected kink states effectively preserve the integrated micro-combs, both QFCs and DKS combs, making them robust against sharp bends and structural imperfections. Our findings show the potential for achieving topologically unidirectional transport of high-dimensional entangled states and phase-locked soliton states in VPC slabs. From an application standpoint, this proposition could open new avenues for high-dimensional information processing utilizing topology in both classical and quantum optics.

\section{Design of VPCs}
\begin{figure*}
\centering
\includegraphics[width=1\textwidth]{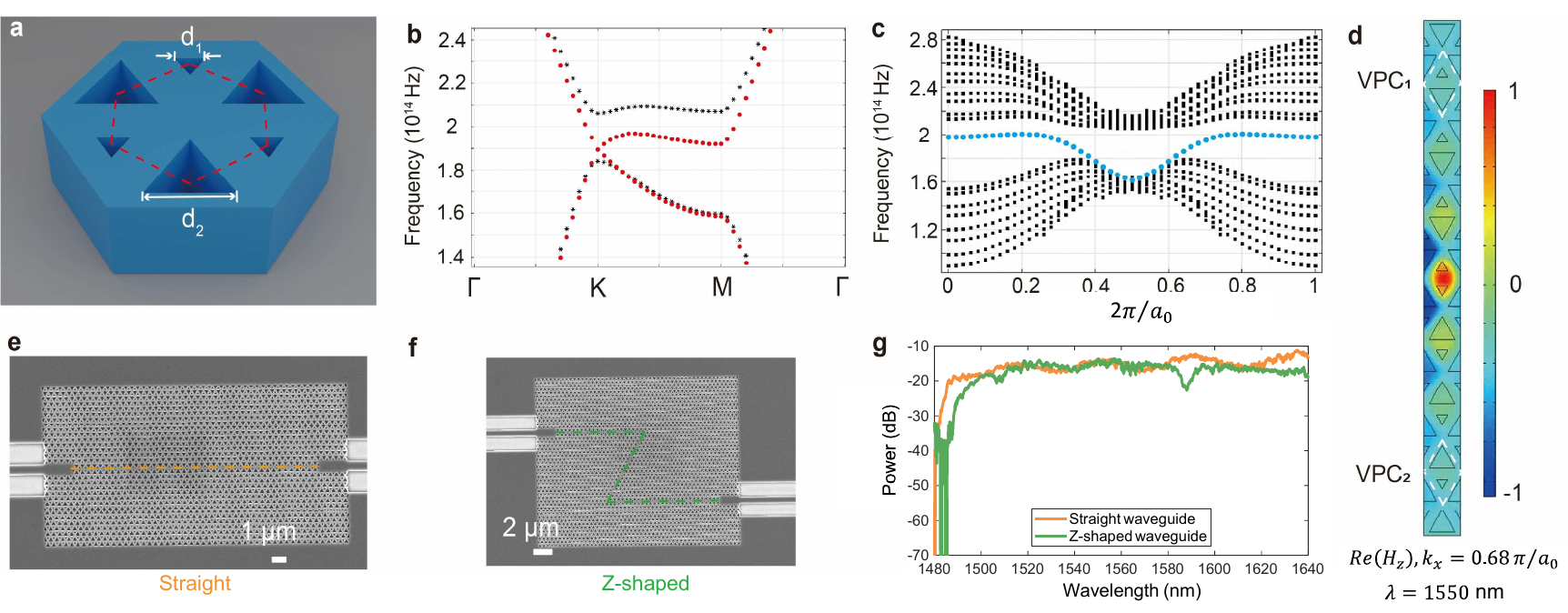}
\caption{Energy band diagram and transmittance of the topological edge state based on VPCs. \textbf{a} The red dashed hexagon in the figure represents the unit cell of the VPC slab with the lattice constant $a_0=433$ nm ($d_1 = 122$ nm and $d_2 = 295$ nm). \textbf{b}  Band diagram of the VPC slab with inversion symmetry (red-dotted curves) compared with inversion symmetry breaking (black-diamond curves). $\Gamma$, K and M denote the high-symmetry points in the first Brillouin zone. \textbf{c}  Edge dispersion of the edge state (blue-diamond curves). \textbf{d} $H_z$ field distributions for the topological edge state based on VPCs along the x direction. \textbf{d,e} Scanning electron microscope (SEM) images of the straight and Z-shaped topological waveguides, respectively. \textbf{f} Measured transmission spectra of the straight interface (orange curve) and Z-shaped interface (green curve), respectively.
}
 \label{fig:1}
\end{figure*}

Our topological VPCs are fabricated based on a SOI  wafer with a thickness of the silicon layer 220 nm (see Methods). 
As shown in Fig.\ref{fig:1}a, the VPCs are composed of a  graphene-like lattice of triangular holes lattice with a lattice constant $a_0=433$ nm.
Two types of triangular stomata with different side lengths form the honeycomb valley photonic structure, where the two side lengths are characterized by $d_1$ and $d_2$, respectively.  

As shown by the red-dotted curve in Fig.\ref{fig:1}b, due to $C_6$ symmetry, 
a Dirac point appears at approximately $\lambda = 1566$ nm at the Brillouin zone corners ($\rm K$ and $\rm K'$) where $d_1 = d_2 = 216$ nm. The inversion symmetry of the VPC can be broken by assigning mismatched wall-length parameters ($d_1 = 122$ nm and $d_2 = 295$ nm), which opens a band gap as illustrated by the black-dotted curve in Fig.\ref{fig:1}b. It is noted that the modes at the $\rm K$ and $\rm K'$ valleys exhibit opposite polarizations, which are commonly referred to as left-handed circular polarization and right-handed circular polarization, respectively.~\cite{3-8,3-9}
Then we define the VPCs as $\rm VPC_1$ ($d_1 = 122$ nm and $d_2 = 295$ nm) and $\rm VPC_2$ ($d_1 = 295$ nm and $d_2 = 122$ nm), respectively. They have the opposite valley Chern numbers, i.e., $C_{v1} < 0$ for $\rm VPC_1$ and $C_{v2} > 0$ for $\rm VPC_2$ (for details, see Supplementary Section I). 
At the interface between $ \rm VPC_1$ and $ \rm VPC_2$, the reversion of the two VPCs leads to the reversion of the sign of the Berry curvature at the valleys, according to the bulk-boundary correspondence, it may exhibit a pair of counter-propagating valley kink states.
As shown in Fig.\ref{fig:1}d, the edge state exists at the topological interface between $\rm VPC_1$ and $\rm VPC_2$, the difference between the valley indices at the K point across the interface is $\Delta C_{_{edge1}}^{K}=C_{_{\nu2}}-C_{_{\nu1}}>0$~\cite{3-8,3-20}, which determines the propagation direction of the fringe state around each valley. 

The dispersion of the edge state is shown in Fig.\ref{fig:1}c (blue curve). There exists a pair of valley kink states with a topological bandgap of valley kink states is approximately 25 THz (from 175 THz to 200 THz). Such a large bandwidth allows the topological transport of OFCs with a bandwidth of approximately 200 nm at telecommunication wavelengths.
This pair of valley kink states with opposite group velocities are locked to valleys, which is referred to as “valley-locked” chirality ~\cite{3-5,3-8}.
Furthermore, to verify the robustness against the sharp turns, we design two types of waveguides, that is, the straight and Z-shaped topological waveguides. The SEM figures of two topological waveguides are shown in Fig.\ref{fig:1}e,f, respectively. The in-and-out coupling of topological waveguides is achieved using lensed fibers with an overall measured insertion loss of 12 dB.
The measured transmission spectra of the light propagation along the interfaces of a 60-degree turn are similar to those along a straight interface, which shows that the propagation of light along the topological waveguide is topologically protected (Fig.\ref{fig:1}g). We can observe a topological bandgap from 1490 nm to 1640 nm, where the transmission (above 1640 nm) can not be measured due to the cutoff excitation wavelengths of the pump laser. Besides, we also simulate the field profiles of valley kink states at the frequency of 193 THz (around 1550 nm), which also reveals the light confinement and robustness against shape corners (see Supplementary Section II for details).

\section{Topological transport of QFCs}
The concept of topological protection harnessed through valley-dependent edge states, finds applicability in the transport of on-chip QFCs. 
We first perform on-chip topological transport of QFCs with frequency entanglement. Our QFC is generated in a $\rm Si_{3}N_{4}$ microring resonator with the high-quality factor ($Q$-factor) of $1.68\times10^{6}$ (see Supplementary Section III for details).

To manipulate the broadband phase matching for spontaneous FWM, we carefully design waveguide cross-section with weak and anomalous group-velocity dispersion (GVD)~\cite{1-1}. As a result of the spontaneous FWM process, a two-photon high-dimensional frequency-entangled state, also referred to as a biphoton frequency comb (BFC), can be accessed~\cite{1-6}. 
For a QFC, an individual signal or idler photon is a coherent superposition of multiple frequency bins. The high-dimensional frequency entanglement of biphoton frequency combs will increase the capability of quantum information processing~\cite{1-7,1-13}.

\begin{figure*}
\centering
\includegraphics[width=1\textwidth]{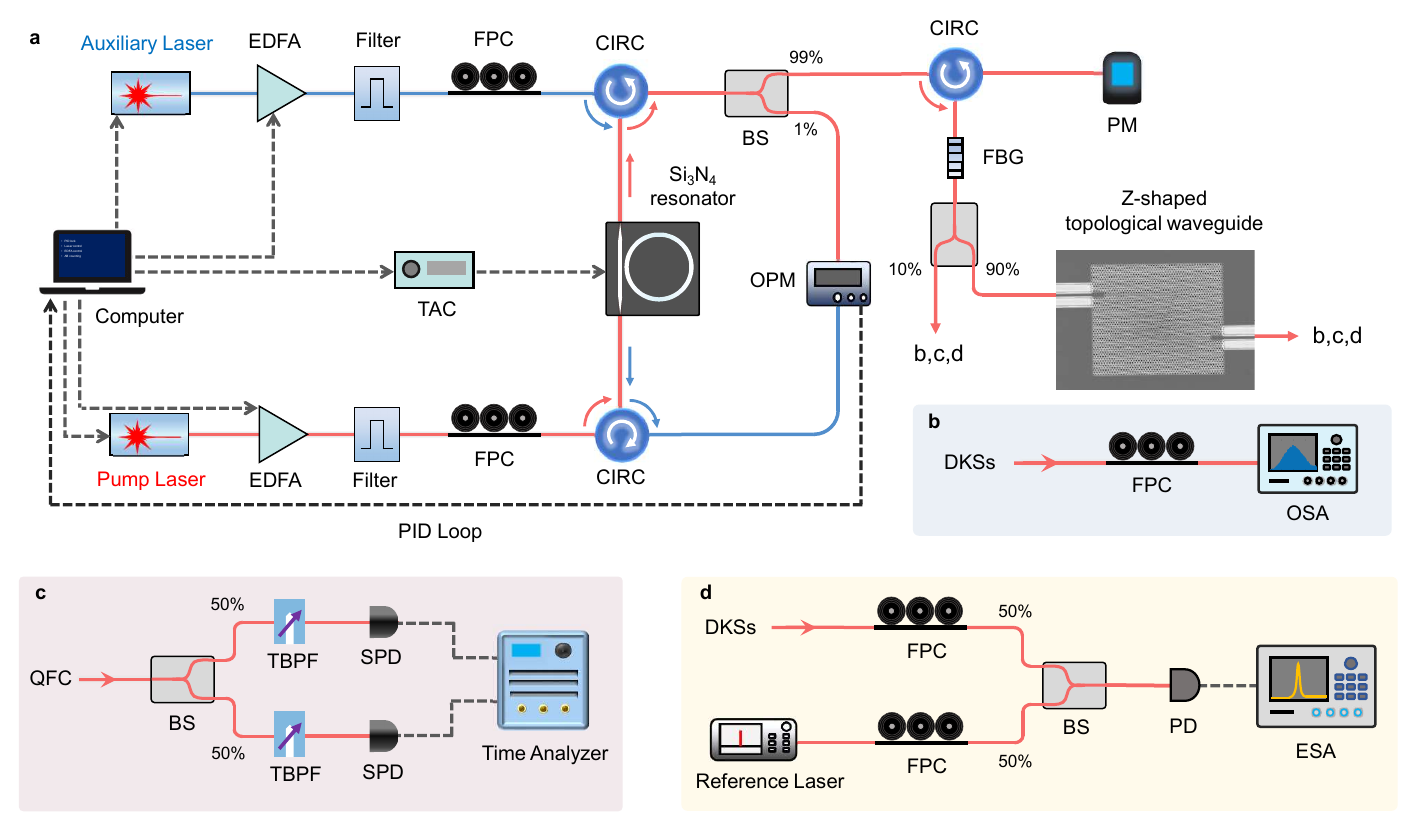}
\caption{\textbf{a} Experimental setup for topological transport of OFCs. For the generation of QFCs, the pump laser is actively tuned by proportional-integral-differential (PID) controller,  while the auxiliary laser is not used. For the generation of DKSs, both two lasers are utilized to pump the resonator. EDFA, erbium doped fiber amplifier; FPC, fiber polarization controller; CIRC, optical circulator; OPM, programable optical power meter; FBG, fiber Bragg grating; OSA, optical spectrum analyzer; ESA, electrical spectrum analyzer. \textbf{b}  Experimental setup for the spectrum of DKSs.
\textbf{c} Characterization of  correlation properties and joint spectral intensity of QFCs. \textbf{d} Experimental setup for RF beatnotes of the single soliton states and a CW reference laser.}
 \label{fig:2}
\end{figure*}

Our experimental setup for the topological transport of quantum frequency combs is schematized in Fig.\ref{fig:2}a (see Methods). We pump the $\rm Si_{3}N_{4}$ micro-resonator to below-threshold pumping at approximately 1550 nm. The spontaneous FWM process leads to a BFC, such a high-dimensional entangled state can be written as~\cite{1-7}:
\begin{equation}
|\Psi\rangle=\sum_{k=1}^N\alpha_k|k,k\rangle_{si},\mathrm{with}\sum|\alpha_k|^2=1,
\label{eq:1}
\end{equation}
where $k$ is the mode number and the complex element $\alpha_k$ represents the amplitude and phase of the signal–idler photon pair, the quantum state $|k,k\rangle_{si}$ is given by
\begin{equation}
|k,k\rangle_{si}=(\Omega-k\Delta\omega)|\omega_{p}+\Omega,\omega_{p}-\Omega\rangle_{si},
\label{eq:2}
\end{equation}
in which $\Omega$ is the frequency deviation from the pump frequency $\omega_{p}$, $\Delta\omega$ is the FSR of the resonator. $f(\Omega)$ is the Lorentzian spectrum function. We can deduce that the quantum state $|k,k\rangle_{si}$ embodies a superposition encompassing numerous symmetrically arranged pairs of signal and idler frequencies around the pump frequency.

\begin{figure*}
\centering
\includegraphics[width=1\textwidth]{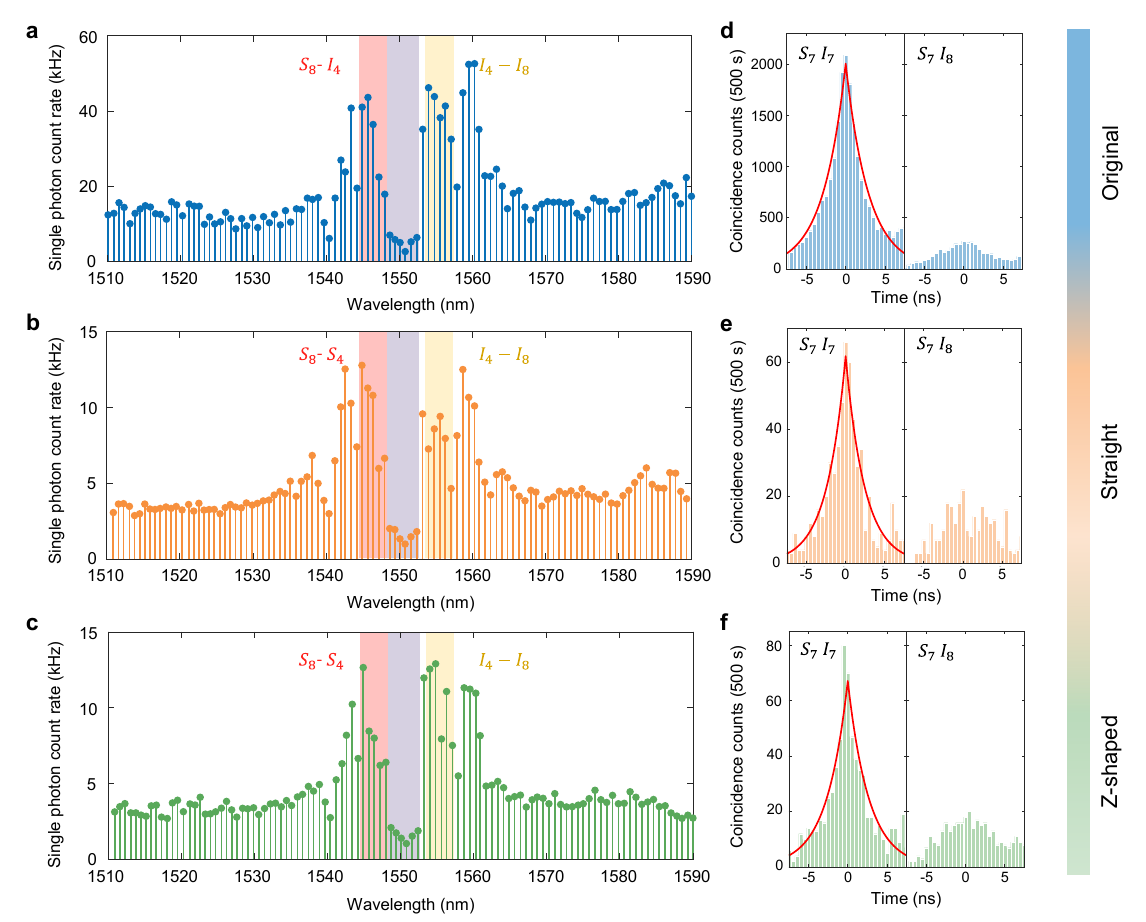}
\caption{Measured single-photon spectrum at the outputs of the \textbf{a} original micro-resonator, \textbf{b}  straight, and \textbf{c} Z-shaped topological waveguides, respectively. The red, blue, and yellow marked regions denote the selected signal modes, idler modes, and several modes eliminated  by the FBG, respectively. Signal-idler coincidence histograms for correlated photon pairs ($S_7I_7$) and uncorrelated photon pairs ($S_7I_8$) at the outputs of the \textbf{d} original micro-resonator, \textbf{e} straight and \textbf{f} Z-shaped topological waveguides, respectively.}
 \label{fig:3}
\end{figure*}

After the rejection of the pump light with a filter, this QFC is coupled into two VPC waveguides. 
Two tunable bandpass filters (TBPFs) are used to select the signal and idler frequency modes, and send them to two InGaAs single photon detectors (SPDs) for counting measurement. 
Figure \ref{fig:3}a depicts the measured single-photon spectrum of the original QFC. Also, the single-photon spectrums of QFCs at the output ports of both straight and Z-shaped topological waveguides are also depicted in Fig.\ref{fig:3}b, c. 
The wavelength bandwidth of these frequency modes is around 0.8 nm, which corresponds to the FSR of $\rm Si_{3}N_{4}$ micro-resonator (about 100 GHz). The single photon spectrum is consistent with the operation bandwidth of valley kink states, while the QFC within the bulk bands is rejected. 
Notably, thanks to the broad bandwidth of topological waveguides, we can clearly observe QFCs ranging from 1510 to 1590 nm, corresponding to around 10 THz bandwidth. As a consequence of energy conservation, the signal and idler photons are highly correlated in frequency and time domains. 

To confirm  the quantum correlation of our QFCs, we record the relative arrival time between correlated photon pairs ($S_7I_7$) and uncorrelated photon pairs ($S_7I_8$) as coincidences. As illustrated in Fig.\ref{fig:3}d-f, the coincidence peaks are only measured with wavelength-correlated cases, while the absence of coincidence peaks between the photon pairs $S_7I_8$ indicates the absence of correlation between frequencies that are not matched. Notably, any event for uncorrelated photon pairs is considered an accidental count, which includes dark counts, background, and uncorrelated photon counts~\cite{1-11}. We fit three coincidence peaks by  Glauber second-order (cross) correlation function $g_{si}(\Delta t)\propto exp[-\Delta t/\tau]$~\cite{3-21}, where $\tau$ is the coherence time of the corrected photons. The fitted coherence times of the photons in three cases are 2.90 ns, 2.69 ns, and 2.45 ns, respectively, which are in agreement with the resonance linewidth ($~115$ MHz). In addition, we characterize the quantum correlation property of photon pairs ($S_7I_7$) by the CAR with 9.7, 10.7, and 10.2, respectively. Note that high on-chip power (6 mW) increases the accidental counts, and then, leads to lower values of CAR. 

\begin{figure*}
\centering
\includegraphics[width=0.9\textwidth]{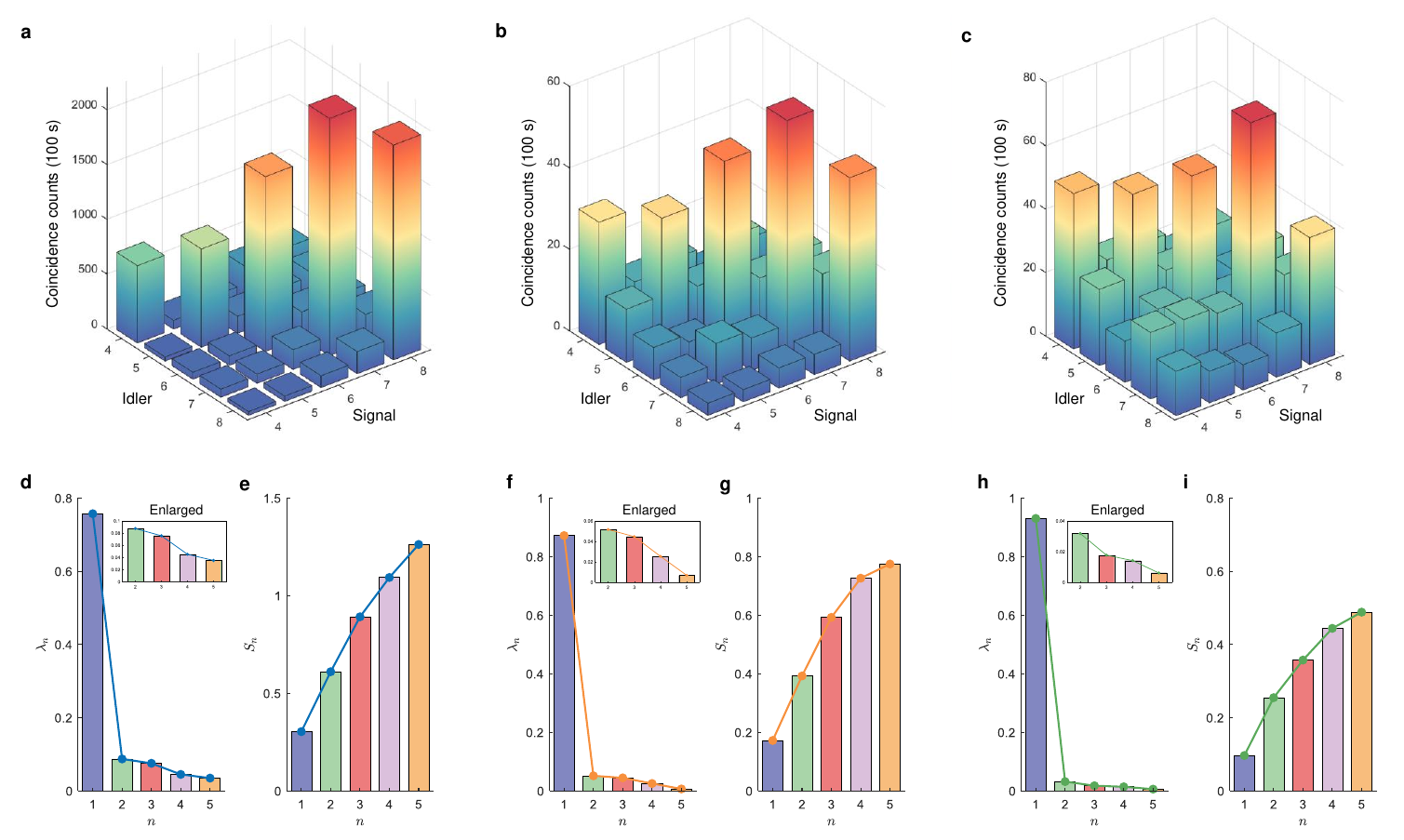}
\caption{Measured joint spectral intensity at the outputs of the \textbf{a} original micro-resonator, \textbf{b}  straight, and \textbf{c} Z-shaped topological waveguides, respectively.
The distributions of normalized Schmidt coefficients $\lambda_n$ for the \textbf{d} original QFC, the QFCs after the transport of the \textbf{f} straight and \textbf{h} Z-shaped topological waveguides, respectively. 
The entanglement entropy $S_k$ for the \textbf{e} original QFC, the QFCs after the transport of the \textbf{g} straight and \textbf{i} Z-shaped topological waveguides, respectively.}
 \label{fig:4}
\end{figure*}

We also performed the calculation of the heralded efficiency $\eta_{h}$, representing the  probability of detecting an idler photon when the signal photon is detected. Generally, the heralded efficiency can be given by $\eta_{h}={cc}/{c_{signal}\eta_{det}}$~\cite{3-23}, where $cc$, $c_{signal}$ are coincidence and signal count rate respectively, $\eta_{det}$ denotes the detection efficiency of the SPD for idler mode (20$\%$). Based on the obtained measurement, we derive a heralding efficiency of $\eta_{h} = 6\%$, without including losses of the experimental setup. 

Furthermore, to evaluate spectro-temporal correlations across the photon-pair spectrum, we conducted measurements of the JSI utilizing two TBPFs. Here we consider time correlation measurement for mode-by-mode photon count covering 5 sideband pairs ($S_{4-8}I_{4-8}$). To mitigate the influence of accidental counting events, we subtract the accidental coincidence counts. It is important to emphasize that the JSI distribution outlines the probability distribution across two dimensions for signal and idler photons~\cite{3-24}.
As shown in Fig.\ref{fig:4}a, photon pairs satisfying the energy conversion relation ($2\omega_p=\omega_s+\omega_i$) exhibit pronounced frequency correlation. In general, the JSI may be given by $|(\mathcal{A}(\omega_s,\omega_i)|^{2}=|A(\omega_s,\omega_i)|^{2}|\Phi(\omega_s,\omega_i)|^{2}$, where $\Phi(\omega_s,\omega_i)$ is corresponding to the phase matching properties and spatial characteristics of the pump beam. Hence, the JSI $|(\mathcal{A}(\omega_s,\omega_i)|^{2}$ demonstrating spectral correlations arises from the energy and momentum conservation~\cite{3-24}.

To quantify the entanglement within quantum frequency combs, a valuable approach known as Schmidt decomposition is employed to assess the separability of the joint spectral amplitude (JSA), with a part of phase information disregarded. The JSA can be approximated from the JSI as $\mathcal{A}(\omega_s,\omega_i)\approx\sqrt{|(\mathcal{A}(\omega_s,\omega_i)|^{2}}$~\cite{1-7}.
It's worth noting that the state $|\Psi\rangle$ is frequency entangled when a probability distribution JSA cannot be decomposed into the function of $\omega_s$ and $\omega_i$. Employing the Schmidt decomposition (see Supplementary Section V), we can access the Schmidt numbers of our QFCs~\cite{3-25}.
Figure \ref{fig:4}d-i show the distributions of normalized Schmidt coefficients $\lambda_n$ and entanglement entropy $S_k$. Note that the Schmit coefficients $\lambda_n$ represent the possibility of obtaining the $n$th quantum state, where the presence of nonzero Schmidt coefficients (greater than 1) indicates the frequency entanglement characteristic of topologically protected QFCs~\cite{3-28}.  Additionally, the entanglement entropy $S_k = -\sum\lambda_n\log_2\lambda_n$ can  describe the degree of entanglement. The entanglement entropy $S_k$ is quantified as 1.41, 0.94, and 0.6, respectively. It serves as evidence of entanglement when $S_k>0$.

The Schmidt number is deducible from  $K=(\sum\lambda_{n}^{2})^{-1}$, yielding the specific values of 1.88, 1.42, and 1.2 for three cases. The small values of these Schmidt numbers arise from undesired accidental counts within off-diagonal components, primarily stemming from afterpulse detection and dark counts of InGaAs SPDs. Notably, these phenomena tend to introduce undesired noise. The Schmidt numbers could be readily enhanced by using superconducting nanowire single photon detectors (SNSPDs), which can significantly reduce the aforementioned issues. 
Assume the off-diagonal elements are entirely eliminated within our measured JSI, a Schmidt number above 4.0 can be attained. 
In a highly multi-dimensional space, the resulting Schmidt number $K$ serves as an effective way for quantifying the high-dimensional frequency entanglement.
The Schmidt number $K$ effectively characterizes the minimum count of pertinent orthogonal modes within a bipartite system.
In the case of a separable system, the JSA satisfies $\mathcal{A}(\omega_s,\omega_i)=\psi_n(\omega_s)\phi_n(\omega_i)$, thereby resulting in a Schmidt number $K=1$~\cite{3-22}. For our QFC, the biphoton entangled quantum state encompasses numerous frequency modes, thereby giving rise to frequency entanglement~\cite{3-25}. 
The unambiguous observation of QFCs serves as clear evidence supporting the achievement of topological transport for high-dimensional quantum entangled states.

Alternatively, one can eliminate the single-photon purity linked to the factorizability of two biphoton states from Schmidt decomposition. Purity is indispensable for achieving high-visibility quantum interferences between photons originating from the same source.
Generally, the single-photon purity can be given by $\text{Tr}(\hat{\rho}_\text{s}^2)$, where $\hat{\rho}_{\mathrm{s}}=\mathrm{Tr}_{\mathrm{i}}(|\Psi\rangle\langle\Psi|)$ is the density operator for the heralded single photon, and $\mathrm{Tr}_{\mathrm{i}}$ denotes the partial trace over the idler mode. The heralded single-photon purity $\text{Tr}(\hat{\rho}_\text{s}^2)$ can be given by $\text{Tr}(\hat{\rho}_\text{s}^2)\overset{}{=}K^{-1}$~\cite{3-24}. Consequently, the single-photon purity for our QFCs in three cases is calculated as 0.53, 0.70, and 0.83, respectively.

\section{Topological transport of DKS combs}
We also perform on-chip topological transport of DKS combs, the experimental procedure is depicted in Fig.\ref{fig:2}a. An integrated $\rm Si_{3}N_{4}$  micro-resonator provides a reliable platform for manipulating temporary DKS combs at the telecommunication wavelengths. 
Before our experimental scheme, we numerically simulated the soliton generation described by Lugiato–Lefever equation (LLE)  (see Supplementary Section VI). 
For solitons generation, it is crucial for the laser wavelength to precisely stabilize at the soliton steps. However, limited by the frequency tunning accuracy or stability of pump lasers, as well as thermal instability within the range of existing solitons, it becomes a challenge for the generation of DKSs within micro-resonators characterized by short thermal lifetimes~\cite{1-1}.
Therefore, a dual-pump method is utilized to produce DKS combs in the same micro-resonator (see Methods), where an  auxiliary laser is introduced to stabilize the intracavity energy, thereby facilitating the extension of soliton steps. 
Except for the benefits of thermal stability, the beating between the auxiliary and pump lasers would be useful for facilitating the generation of soliton crystals~\cite{3-26}.

\begin{figure*}
\centering
\includegraphics[width=1\textwidth]{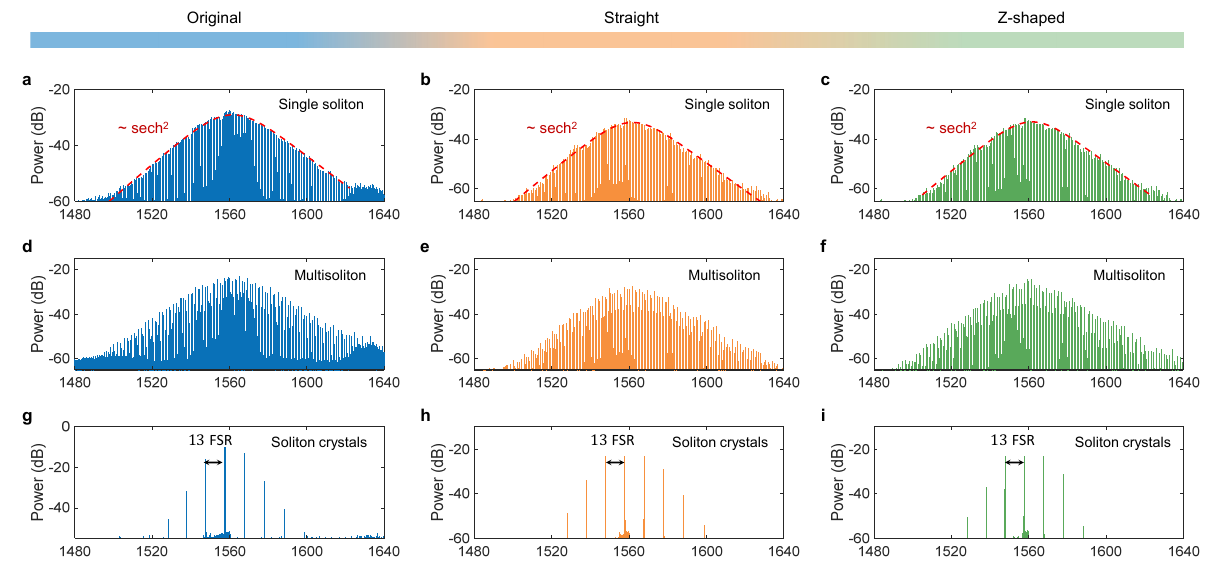}
\caption{Optical spectrum of the single soliton state at the outputs of a the original micro-resonator, \textbf{b}  straight, and \textbf{c} Z-shaped topological waveguides, respectively. 
Optical spectrum of the multisoliton state at the outputs of \textbf{d} the original micro-resonator, \textbf{e} straight, and \textbf{f} Z-shaped topological waveguides, respectively. 
Optical spectrum of perfect soliton crystals at the outputs of \textbf{g} the original micro-resonator, \textbf{h} straight, and \textbf{i} Z-shaped topological waveguides, respectively.}
 \label{fig:5}
\end{figure*}

With the scanning of the pump laser from ‘blue’ to ‘red’ laser-cavity, one can clearly observe the comb evolution processes including Turing rolls, chaotic states (breathing solitons), multisolitons, and single solitons, respectively~\cite{3-27}.
Generally, a sharp drop in the generated comb power signifies the arrive of the bistable region, precisely revealing the appearance of solitons, which is also commonly referred to as the 'soliton step'~\cite{1-1}.

The produced DKS combs are split by a 90:10 beam splitter (BS), where 10$\%$ of the resulting comb output is directed towards an OSA to assess the spectrum of the original Kerr combs. Meanwhile, the remaining 90$\%$ of the output is coupled into the topological waveguide following polarization control. Subsequently, the output of the topological waveguide is directed to another OSA, allowing for the monitoring of its spectral properties.

DKS states are readily accessed by stopping the laser at distinct steps. A single soliton state and its transmission spectrum at the output ports of both straight and Z-shaped topological waveguides are shown in Fig.\ref{fig:5}a-c. Note that the frequency modes of the pump laser and FWM sidebands generated from the auxiliary laser in the optical spectrum are removed.
The associated optical spectrum reveals line spacing matches the resonator FSR over a range of around 20 THz, and also exhibits smooth $\rm sech^2$-shaped spectral envelopes. The single soliton spectrum has a 3-dB bandwidth of $\sim$29.3 nm corresponding to a soliton pulse width of $\sim$87.5 fs. 
To further verify the coherence of single soliton states, we beat one of the spectral lines of the comb with a CW laser. The beat notes show distinct frequency lines with resolution bandwidths (RBWs) of 100 kHz and signal-to-noise ratios of 30 dB (see Supplementary Section VII), leading to outstanding low-noise characteristics.

Due to the inherently stochastic nature of intracavity dynamics, multisolitons with a random soliton number $N$ could be assessed during the frequency-tuning procedure~\cite{1-1}. The spectral envelopes of the multisoliton state emerge from the interference between multiple solitons, and become more complicated as the number of solitons increases~\cite{3-27}. In other words, the intracavity solution for the multisoliton state can be expressed as a summation of distinct, independent soliton solutions situated at various positions, as depicted in Figure \ref{fig:5}d-f. Strikingly, An adiabatic backward tuning method can successfully reduce the soliton number, which makes it reliable to access the single soliton state from the multisoliton state~\cite{1-1}.

We also access perfect soliton crystals using the forward tuning pump at relatively low pump powers (100 mW). Perfect soliton crystals, that is, a collective self-organization result of the multiple solitons. The transmission spectrum of original perfect soliton crystals is shown in Fig.\ref{fig:5}g, while the spectra after the transport of both straight and Z-shaped topological waveguides are shown in Fig.\ref{fig:5}h,i. The combs show several evenly distributed supermodes separated by 13 FSRs, leading to a mode spacing of 1.24 THz. 
Also, our perfect soliton crystals can be regarded as a single soliton  with  larger FSR, which can be applied in the generation of ultra-high-repetition
rate DKS combs~\cite{3-28}. 
In the time domain, there are 13 DKSs in the resonator with a time-independent pulse separation of $2\pi /13$, reaching the maximum allowed number of solitons existing with the given pump. 
Since the comb power is distributed in such 13 supermodes, the power of each supermode is enhanced by $13^2$ times, and energy conversion efficiency by 13 times compared with
the single soliton state~\cite{3-27}.

Due to the broad bandwidth of topological waveguides, we show the topological transport of DKS combs with a range of around 20 THz, and all DKS combs can smoothly pass through sharp bends without visible loss. Despite transport along sharp turns, our DKSs retain their characteristics, including the specific spectral envelopes and low-noise operation.

\section{Conclusion}
In this work, we  experimentally demonstrated on-chip topological transport for optical frequency combs at telecommunication wavelengths. We show topological valley kink states are robust against sharp turns in a VPC slab. We access both the two-photon frequency-entangled QFC and mode-locking DKS combs with a $\rm Si_{3}N_{4}$ micro-resonator. For our QFCs, a coherent superposition of multiple frequency modes leads to a frequency-entangled qudit state. This qudit state maintains its strong frequency correlation with the transport along topological interfaces with sharp bends.  In particular, the QFCs also have the potential to be energy-time entanglement, which can be detected by quantum interference in a Franson-type interferometer~\cite{1-11,1-12}. Such topologically protected QFCs provide robust, complex, high-controllable, and large-scale quantum resources, which may scale up quantum communication, as well as quantum information processing. In addition, we also demonstrate the topological transport of mode-locked DKS combs in our design, revealing that DKSs keep their perfect spectral phase and temporal pulse shape. The ultrahigh stability and low-noise characteristics of topologically protected DKS combs provide new applications in metrology and spectroscopy.

\section{Methods}

\subsection{Device fabrication}

We fabricated the proposed topological device on an SOI wafer with a 220 nm thick top silicon layer and  3 $\mu m$ thick buried silicon layer. 
The edge coupler, silicon waveguide, and VPC structures were etched to a depth of 220 nm. 
Subsequently, we deposited a 1 $\mu m$ thick $\rm SiO_2$ cladding using plasma-enhanced chemical vapor deposition (PECVD). Finally, the whole chip was deeply etched and cut into several individual chips. The devices were fabricated using electron beam lithography (Vistec EBPG 5200+) and an inductively coupled plasma etching process (SPTS DRIE-I). For details on the fabrication process of the structures presented in this study, see Refs. ~\cite{4-1,4-2}.

\subsection{Experimental setup for QFCs generation}
In this experiment, we use a compact CW laser (Pure Photonics) with a wavelength of 1550.78 nm to pump the $\rm Si_{3}N_{4}$ micro-resonator and an EDFA to amplify the pump. Two cascaded bandpass filters are used to suppress the ASE noise produced by the EDFA. The measured power coupling into the resonator is 6 mW. A TEC is applied to control the thermal instability of the resonator with the injection of the pump. The output QFC is coupled into topological waveguides after the polarization adjustment by a FPC. At the output port, a 99:1 BS is connected, where 1$\%$ of the output power is detected by a programable OPM (Joinwit) to monitor the pump power. This part of the output pump monitored by an OPM is used to actively control the wavelength of the CW laser by PID algorithm. The application of PID control ensures a stable resonant state for long-time coincidence measurement. The other 99$\%$  of output power is rejected by an FBG, and the remaining QFCs are split by a 50:50 BS. The signal and idler modes are selected by two TBPFs (WL Photonics) with a bandwidth of 0.11 nm, and detected by two InGaAs SPDs (Aurea Technology). 
For the single photon spectrum measurement, the SPDs are set to gated mode with 20$\%$  quantum efficiency and 10 $\rm \mu$s dead time. For the JSI measurement, the SPDs are set to gated mode (20 MHz repetition rate) with 20$\%$ quantum efficiency and 10 $\rm \mu$s dead time and then sent to a time analyzer to record coincidence events between signal and idler photons.

\subsection{Experimental setup for DKS combs generation }
In this experiment, we use the dual-pump method to produce DKS combs by a computer-controlled soliton generation system~\cite{4-3}. The pump laser is excited by the same CW laser, and amplified to 0.4 W by an EDFA. The EDFA amplified spontaneous emission (ASE) noise is suppressed by two cascaded bandpass filters. An extra auxiliary laser with 1 W power is employed to stabilize the intracavity energy for extending the steps of solitons. To reduce the crossed interaction of the two pumps, they are changed to orthogonal modes by two independent fiber FPCs. Then these two lasers are injected into the bus waveguide in opposite directions, with two circulators separating the input and residual pump light. Besides, a TEC is applied to control the thermal instability of the micro-resonator. The output comb is split by a 99:1 BS, with 1$\%$ of the output power detected by an OPM. The other 99$\%$ of output power is separated by a 90:10 BS, with 10$\%$ of residual output sent to an OSA to measure the spectrum of generated Kerr combs, while 90$\%$ of residual output is coupled into the topological waveguide after the polarization control. Then the output of the topological waveguide is sent to another OSA to monitor its spectrum.
For the computer-controlled soliton generation system, the auxiliary laser is tuned at the resonance wavelength to produce primary FWM sidebands. The pump laser is automatically tuned from the blue to red side to assess the soliton states. in this process, the script will assess the intracavity state by the measured output power. Once the soliton state is accessed, the automation script gives the “stop” order for the pump laser (see Supplementary Section IV). In this case, the soliton state will hold for several hours. 

\begin{acknowledgments}
This work is supported by the Key-Area Research and Development Program of Guangdong Province (2018B030325002), the National Natural Science Foundation of China (62075129, 61975119, 62305210), the SJTU Pinghu Institute of Intelligent Optoelectronics (2022SPIOE204), and the Science and Technology on Metrology and Calibration Laboratory (JLJK2022001B002).
\end{acknowledgments}




%

\end{document}


\renewcommand{\thefigure}{S\arabic{figure}}
\renewcommand{\theequation}{S\arabic{equation}}
\renewcommand{\thetable}{S\arabic{table}}

\preprint{APS}

\title{Supplementary Materials: On-chip topological transport of optical frequency combs in silicon-based valley photonic crystals}

\author{Zhen Jiang$^{1,2,\#}$}%
\author{Hongwei Wang$^{1,\#}$}%
\author{Yuechen Yang$^{1,2}$}%
\author{Yang Shen$^{1,2}$}%
\author{Bo Ji$^{1,2}$}%
\author{Yanghe Chen$^{1,2}$}%
\author{Yong Zhang,$^{1}$}%
\author{Lu Sun,$^{1}$}%
\author{Zheng Wang,$^{1}$}%
\author{Chun Jiang$^{1}$}%
\email{cjiang@sjtu.edu.cn}	
\author{Yikai Su$^{1}$}%
\email{yikaisu@sjtu.edu.cn}	
\author{Guangqiang He$^{1,2}$}%
\email{gqhe@sjtu.edu.cn}	
\affiliation{%
 $^1$State Key Laboratory of Advanced Optical Communication Systems and Networks, Department of Electronic Engineering, Shanghai Jiao Tong University, Shanghai 200240, China\\
 $^2$SJTU Pinghu Institute of Intelligent Optoelectronics, Department of Electronic Engineering, Shanghai Jiao Tong University, Shanghai 200240, China \\
 $^\#$These authors contributed equally to this work.
}%

\maketitle

\section{ Theory of valley topological photonics (VPCs)}
For unperturbed unit cells exhibiting $C_6$ lattice symmetry, degenerate Dirac points are present in the $\rm K$ and $\rm K'$ valleys. The effective Hamiltonian in the vicinity of the $\rm K$ ($\rm K'$) point is given by~\cite{s1, s2, s3}.

\begin{equation}
H_{K/K^{\prime}}=\tau_z\nu_D(\sigma_x\delta k_x+\sigma_y\delta k_y),
\label{eq:S1}
\end{equation}	
where $v_{D}$ is the group velocity, $\sigma_{x}$ and $\sigma_{y}$ are the Pauli matrices, $\delta\vec{k}=\vec{k}-\vec{k}_{K/K'}$ denotes the deviation of the wavevector. With the distortion of the unit cell ($d_1 \neq d_2$), the Hamiltonian can be rewritten as 
\begin{equation}
H_{K/K^{\prime}}=\tau_z\nu_D(\sigma_x\delta k_x+\sigma_y\delta k_y)+\tau_z\gamma\sigma_z,
\label{eq:S2}
\end{equation}	
where $\tau_z=1(-1)$ denotes the $\rm K$ ($\rm K'$) valley pseudospin, $\sigma_{x,y,z}$ denotes the Pauli matrices, $\nu_{_D}$ is the group velocity, $\gamma$ is the strength of the symmetry-breaking perturbation. The perturbation is $\gamma_{1}\propto\left[\int_{B}\varepsilon_{z}ds-\int_{A}\varepsilon_{z}ds\right]$ ($\rm VPC_1$) and $\gamma_2\propto\left[\int_D\boldsymbol{\varepsilon}_zds-\int_C\boldsymbol{\varepsilon}_zds\right]$ ($\rm VPC_2$), respectively, where $\int\varepsilon_zds$ is the integration of dielectric constant $\varepsilon_z$ at the positions of $A$ and $B$, respectively.
In this context, the VPCs are $d_{A}=0.72a_0$ and $d_{B}=0.28a_0$, consequently leading to the inequality $\int_B\varepsilon_zds<\int_A\varepsilon_zds$. Additionally, we can get $|\gamma_1|>|\gamma_2|$.
This indicates that the modes at the $\rm K$ and $\rm K'$ valleys exhibit opposite circular polarizations, specifically, left-handed circular polarization (LCP) and right-handed circular polarization (RCP).
The valley Chern numbers of VPCs can be given by~\cite{s1,s4}
\begin{equation}
C_{K/K^{\prime}}=\frac{1}{2\pi}\int_{HBZ}\Omega_{K/K^{\prime}}(\delta\vec{k})dS=\pm1/2,
\label{eq:S3}
\end{equation}	
where $\Omega=\nabla_k\times\vec{A}(k)$ is the Berry curvature, $\vec{A}(k)$ is the Berry connection,
and this integration region contains half of the Brillouin zone. 
Hence, the disparity in the valley Chern numbers of the system is computed as $|C_{K/K'}|=1$, affirming the topological characteristics of VPCs.

\section{Numerical simulation of topological edge states}
\begin{figure*}
\centering
\includegraphics[width=1\textwidth]{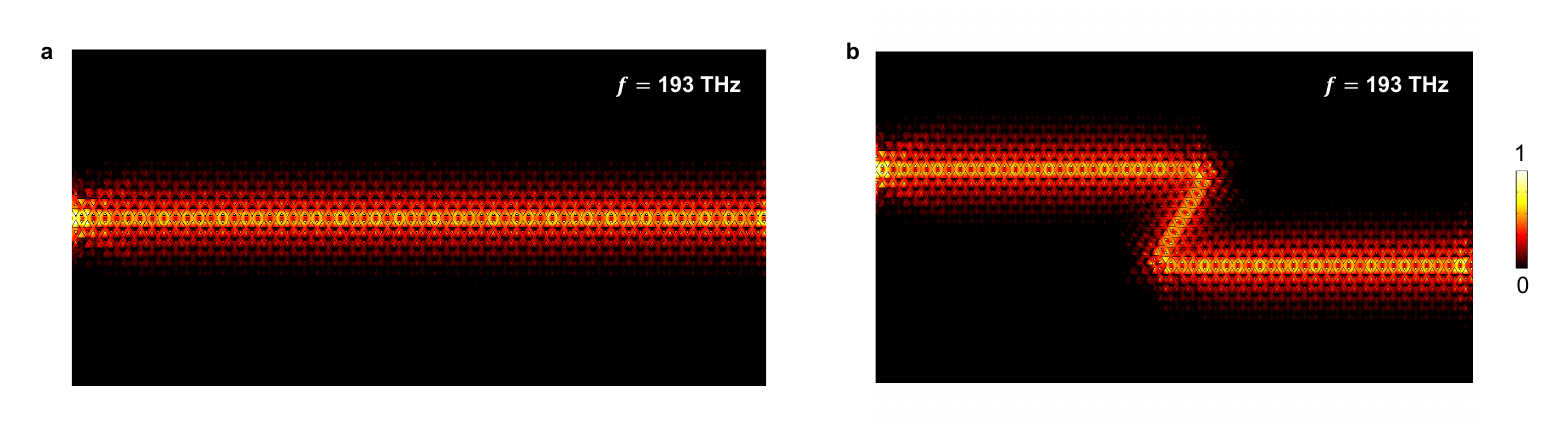}
\caption{Simulated field profiles of the valley kink states at the frequency of 193 THz (around 1550 nm) at different interfaces.}
 \label{fig:S1}
\end{figure*}

Topologically protected edge states in VPCs, also known as valley kink states~\cite{s5}, can be observed at the boundary between $\rm VPC_1$ and $\rm VPC_2$. To confirm the robustness of these edge states, we design a straight waveguide and a Z-shaped topological waveguide. The simulated field profiles of the valley kink states at the frequency of 193 THz (around 1550 nm) are shown in Fig.~\ref{fig:S1}. The results reveal that valley kink states are highly centralized at the interfaces, and show robustness against shape corners.

\section{Micro-resonator Characterization}

\begin{figure*}[h]
\centering
\includegraphics[width=1\textwidth]{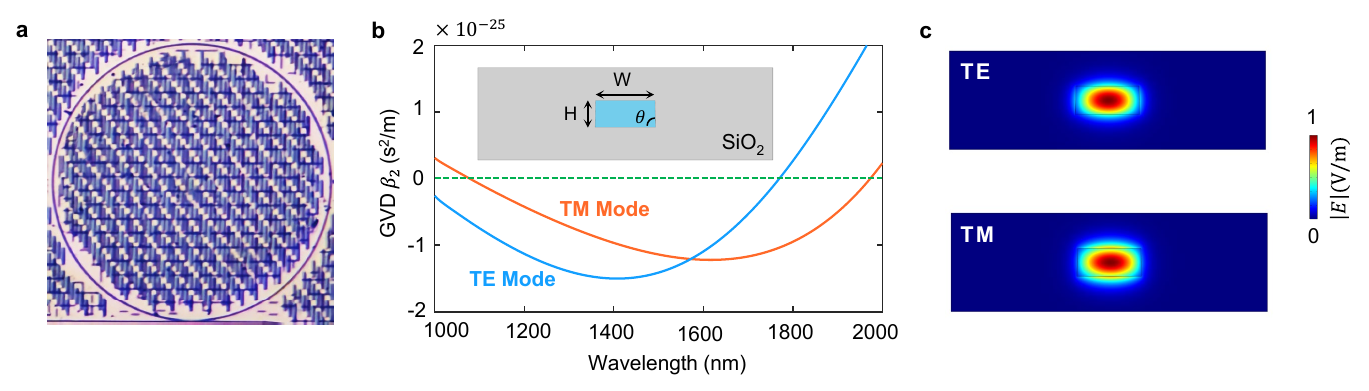}
\caption{\textbf{a} Microscopy image of the  $\rm Si_{3}N_{4}$  micro-ring with $W=1.8$ $\rm\mu m$, $H=0.8$  $\mu m$, $\theta=89^{\circ}$, and gap width of  0.45 $\mu m$. \textbf{b} Simulated GVD curve for TE and TM modes respectively, where the inset denotes a diagram of the waveguide cross-section. \textbf{c} Simulated mode profiles of TE and TM modes, respectively.}
 \label{fig:S2}
\end{figure*}

We utilize a $\rm Si_{3}N_{4}$ micro-resonator to generate both quantum frequency combs (QFCs) and dissipative Kerr soliton (DKS) combs. Consequently, dispersion engineering becomes a critical manipulation for producing predictive combs with the resonator. The waveguide cross-section is numerically simulated using the COMSOL Multiphysics software. In this context, we select a waveguide cross-section with $W=1.8$ $\rm\mu m$, $H=0.8$  $\mu m$, $\theta=89^{\circ}$. A schematic of the waveguide structure is presented in the inset of  Fig.~\ref{fig:S2}b. The micro-resonator is coupled by a bus waveguide with a gap of 0.45 $\mu m$.

To achieve the generation of both QFCs and DKS combs, it is necessary to satisfy the condition of group-velocity dispersion (GVD) with an anomalous feature. By expanding the propagation phase constant $\beta$ in a Taylor series, we can estimate the second term $\beta_2$ by:
\begin{equation}
\beta_2=\frac{1}{c}\bigg(2n\frac{dn(\omega)}{d\omega}+\omega\frac{d^2n(\omega)}{d\omega^2}\bigg),
\label{eq:S4}
\end{equation}	
where $n(\omega)$ signifies the effective refractive index. Figure~\ref{fig:S2}b illustrates the simulated GVD curve for fundamental modes with transverse-electric (TE) and transverse-magnetic (TM) polarizations respectively. It suggests a near-zero anomalous GVD around the wavelength of 1550 nm. Additionally, the corresponding mode profiles of TE and TM modes are depicted in Fig.~\ref{fig:S2}c.
Following a meticulous design of the $\rm Si_{3}N_{4}$ micro-resonator, we entrust the LIGETEC to fabricate resonators with the AN800 technology. A microscopy image of the $\rm Si_{3}N_{4}$ micro-ring is presented in Fig.~\ref{fig:S2}a.

\begin{figure*}[h]
\centering
\includegraphics[width=0.8\textwidth]{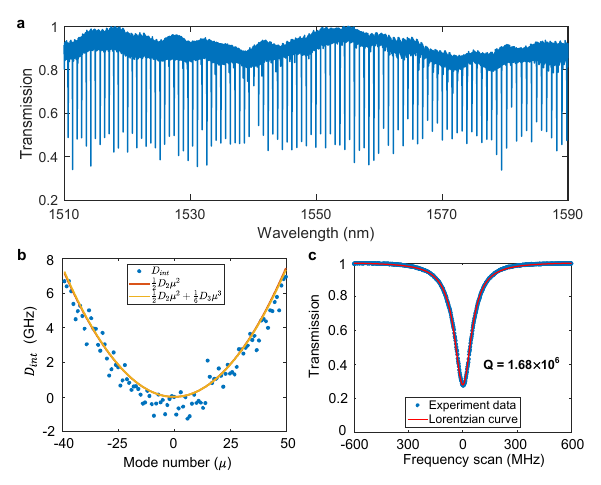}
\caption{\textbf{a} Measured transmission spectrum of the $\rm Si_{3}N_{4}$ micro-resonator. \textbf{b} Dispersions of the micro-resonator extracted from the measured transmission. \textbf{c} Lorentzian fitting of the resonant dip, revealing the $Q$-factors of  $Q=1.68\times10^{6}$, $Q_{in}=2.20\times10^{6}$, and $Q_{ex}=7.10\times10^{6}$, respectively.}
 \label{fig:S3}
\end{figure*}

To measure the transmission spectrum of the micro-resonator, a tunable semiconductor laser (TSL, Santec) is swept from 1510 nm to 1590 nm. The result is illustrated in Fig.~\ref{fig:S3}a, which leads to a free spectral range (FSR) of 95.75 GHz. The Taylor expansion for the cavity resonant mode $\omega_\mu$ is expressed by 
\begin{equation}
\omega_\mu=\omega_0+\sum_{n=1}^{\infty}{D_n\frac{\mu^n}{n!}},
\label{eq:s5}
\end{equation}
where $\mu (\mu\in{\mathbb Z})$ denotes the mode indexing, $\omega_0$ corresponds to the pumped resonant mode. The expansion terms (also referred to as the high-order dispersion) can be written as $D_n=d^n\omega_\mu/d\mu^n$ at $\omega=\omega_{0}$. 
Another term so-called integrated dispersion is used to describe the resonator properties, which is given by $D_{int}(\mu)=\omega_\mu-\left(\omega_0+\mu D_1\right)$. Figure~\ref{fig:S3}b shows dispersions of the micro-resonator extracted from the measured transmission, indicating a value of the second-order dispersion $D_2=5.95\times10^6$ rad/s. 
The appearance of resulting OFCs (both classic and quantum) depends on the quality factor ($Q$-factor) of micro-resonators. 
The total $Q$-factor can be expressed as $1/Q = 1/Q_{in} + 1/Q_{ex}$, in which $Q_{in}$ and $Q_{ex}$ are intrinsic quality factor and external quality factor, respectively. Here we sweep a resonant dip with a narrower linewidth at the pump wavelength.
The Lorentzian fitting of the resonant dip shown in Fig.~\ref{fig:S3}c  reveals $Q_{in}=2.20\times10^{6}$ and $Q_{ex}=7.10\times10^{6}$. Hence, the total quality of the micro-resonator is calculated to be $Q=1.68\times10^{6}$. And the loss can be calculated as $\kappa=7.22\times10^{8}$ ${rad/s}$, $\kappa_{in}=5.51\times10^{8}$ ${rad/s}$, and $\kappa_{ex}=1.71\times10^{8}$ ${rad/s}$, respectively. Consequently, the coupling efficiency is determined as $\eta=\kappa_{ex}/(\kappa_{ex}+\kappa_{in})=0.24$. This value indicates that the $\rm Si_{3}N_{4}$ micro-ring, featuring a gap of 0.45 $\mu m$, corresponds to an under-coupling scenario~\cite{s5}.

\section{Prototype of optical frequency combs}

\begin{figure*}[h]
\centering
\includegraphics[width=1\textwidth]{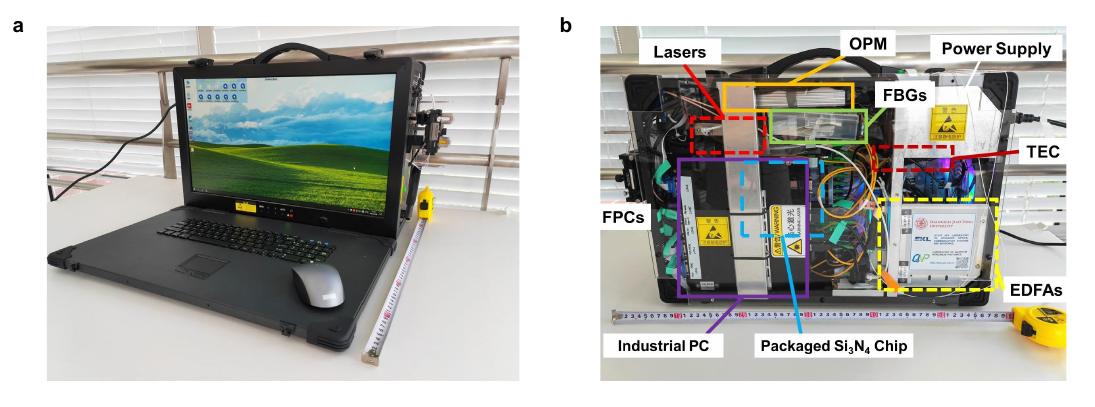}
\caption{A prototype for the generation of both QFCs and DKS combs.}
 \label{fig:S4}
\end{figure*}

To downsize the generation system for both QFCs and DKS combs, we have advanced our previous prototype~\cite{s6} to establish compatibility for these applications. 
We focus on building a standalone microcomb source integrating all the necessary hardware into a 21.5-inch chassis, to reduce possible off-board destabilization during the operation processing. Fig.~\ref{fig:S4} shows our promoted microcomb generation prototype. Compared with our previous system, in this iteration, we optimize the generation setup for producing DKS combs with the dual-pump method, and QFCs with the single-pump excitation method. A TEC with a feedback controller is packaged at the bottom of $\rm Si_{3}N_{4}$ chip, the precision of this temperature stabilization subsystem is 2 mK. Besides, we improve the concentration and stability of optical and electric circuit arrangement.

For the software of the prototype, we write corresponding scripts to produce DKS combs and QFCs respectively. In the case of DKSs generation, the auxiliary laser is controlled to reach a resonator mode far away from the pump resonator wavelength automatically. The pump laser is tuned from the blue to red side, and stops until the automation script discerns the soliton states according to the intracavity power. In the case of QFC generation, the frequency correlation measurement needs a stable resonant situation. To ensure this, we use a proportional-integral-differential (PID) loop to actively control the wavelength of the pump laser. In this script, once the pump laser accesses the resonant dip, the PID loop activates and dynamically maintains the stabilization of the output power.

\section{Theoretical analysis of quantum frequency combs}
Here we theoretically discuss the generation of QFCs in our $\rm Si_{3}N_{4}$ resonators. The biphoton states are generated from spontaneous FWM process, satisfying $2\omega_{p}=\omega_{s}+\omega_{i}$ and $2\vec{k}_{p}=\vec{k}_{s}+\vec{k}_{i}$, where $\omega_{p,s,i}$ and $\vec{k}_{p,s,i}$ are the frequencies and wavevectors of four photons. The nonlinear Hamiltonian of the FWM process in the resonator can be given by
\begin{equation}
\begin{aligned}H_{non}&=\frac{\chi^{(3)}}{2L}\int_{-L}^0dzE_{p}^{(+)}E_{p}^{(+)}E_{s}^{(-)}E_{i}^{(-)}+h.c.,\end{aligned}
\label{eq:s6}
\end{equation}
where $L$ is the cavity length, $h.c.$ denotes Hermitian conjugate. The pump field takes the form of classical wave~\cite{s7}:
\begin{equation}
E_{{p}}^{(+)}(z,t)=E_{{p}}e^{i[k_{{p}}z-\omega_{{p}}t]}e^{-i\Gamma Pz},
\label{eq:s7}
\end{equation}
where the  term $e^{-i\Gamma Pz}$ represents the pump self-phase modulation, $\Gamma$ is the nonlinear parameter of $\rm Si_{3}N_{4}$, $P$ is the intracavity power. In addition, the quantized field of signal and idler modes can be given by~\cite{s7}
\begin{equation}
\begin{array}{rcl}E_{s,i}^{(-)}(z,t)&=&\sqrt{\frac{\hbar\omega_{s,i}}{2\varepsilon_0n_{s,i}cA_{eff}}}\frac{\sqrt{\gamma_{s,i}\Delta\omega}}{2\pi}\sum\limits_{\mu}\int_{-\infty}^{\infty}d\Omega_{s,i}\frac{a_{s,i}^\dagger(\omega_{\mu_s,\mu_i}+\Omega_{s,i})}{\gamma_{s,i}/2-i\Omega_{s,i}}e^{-i[k_{s,i}z-(\omega_{\mu_s,\mu_i}+\Omega_{s,i})t]}.\end{array}
\label{eq:s8}
\end{equation}
where $\Delta\omega$ is the free spectral range (FSR), $A_{eff}$ is effective field cross-section area, $\gamma_{s,i}$ is the linewidth of the cavity, $\omega_{\mu_s,\mu_i}$ is the $\mu$-th central frequency with $\omega_{\mu_s,\mu_i}=\omega_p\pm\mu\Delta\omega$, and $\Omega_{s,i}$ denotes the deviation from $\omega_{\mu_s,\mu_i}$.
Therefore, the nonlinear Hamiltonian can be given by 
\begin{equation}
\begin{array}{rcl}H_{non}&=& \hbar\eta\sum\limits_{\mu_s}\sum\limits_{\mu_i}\int_{-\infty}^{\infty}d\Omega_s\int_{-\infty}^{\infty}d\Omega_i
\frac{\sqrt{\gamma_s\gamma_i}F_{\mu_s,\mu_i}}{(\gamma_s/2-i\Omega_s)(\gamma_i/2-i\Omega_i)} a_{s}^\dagger(\omega_{\mu_s}+\Omega_s)a_{i}^\dagger(\omega_{\mu_i}+\Omega_i)e^{i[(\mu_s+\mu_i)\Delta\omega+\Omega_s+\Omega_i]t} +h.c..\end{array}
\label{eq:s9}
\end{equation}
Here, the constant term is
\begin{equation}
\eta=\frac{{E_P}^2}{16\pi^2\epsilon_0cA_{eff}}\sqrt{\frac{\omega_s\omega_i}{n_sn_i}}\chi^{(3)}\Delta\omega.
\label{eq:s10}
\end{equation}
The interaction between signal and idler modes is 
\begin{equation}
F_{\mu_s,\mu_i}=\int_{-L}^0dz e^{i(2k_p-k_s-k_i-2\Gamma P)}, 
\label{eq:s11}
\end{equation}
where we can define the phase-matching condition term $\Delta k=2k_p-k_s-k_i-2\Gamma P$. Applying the  first-order perturbation theory, the biphoton state can be calculated by 
\begin{equation}
|\Psi\rangle=\frac{1}{i\hbar}\int_{-\infty}^{\infty}dtH_{non}|0\rangle 
\label{eq:s12}
\end{equation}
Using the Eq.~\ref{eq:s9}, we can rewrite the biphoton state as~\cite{s8}
\begin{equation}
\begin{aligned}
|\Psi\rangle
&=\eta \sum_{\mu_s}\sum_{\mu_i}\int_{-\infty}^{\infty}d\Omega_s\int_{-\infty}^{\infty}d\Omega_{i}\frac{\sqrt{\gamma_s\gamma_i}}{(\gamma_s/2-i\Omega_s)(\gamma_i/2-i\Omega_i)}\\
&\times e^{(i\Delta k L)}a_{s}^\dagger(\omega_{\mu_s}+\Omega_s)a_{i}^\dagger(\omega_{\mu_i}+\Omega_i)|0\rangle\text{sinc}{(\Delta k L)},\end{aligned}.
\label{eq:s13}
\end{equation}

In our resonators, the frequency deviation is much smaller than the FSR, that is, |$\Omega_{{s,i}}|\ll\Delta\omega_{{s,i}}$, and the index numbers of signal and idler modes are same. Therefore we can get $\mu_s=\mu_i=\mu$ and $\Omega_s=-\Omega_i=\Omega$. 
Using these reasonable assumptions, the biphoton state can be simplified as
\begin{equation}
\begin{array}{rcl}|
\Psi\rangle&=&\eta \sum\limits_{\mu}\int_{-\infty}^{\infty}d\Omega\frac{\sqrt{\gamma_s\gamma_i}e^{(i\Delta k L)}}{(\gamma_s/2-i\Omega)(\gamma_i/2+i\Omega)}\text{sinc}{(\Delta k L)}a_{s}^\dagger(\omega_{p}-m\Delta\omega+\Omega)a_{i}^\dagger(\omega_{p}+m\Delta\omega-\Omega)|0\rangle.\end{array}
\label{eq:s14}
\end{equation}

\begin{figure*}[h]
\centering
\includegraphics[width=1\textwidth]{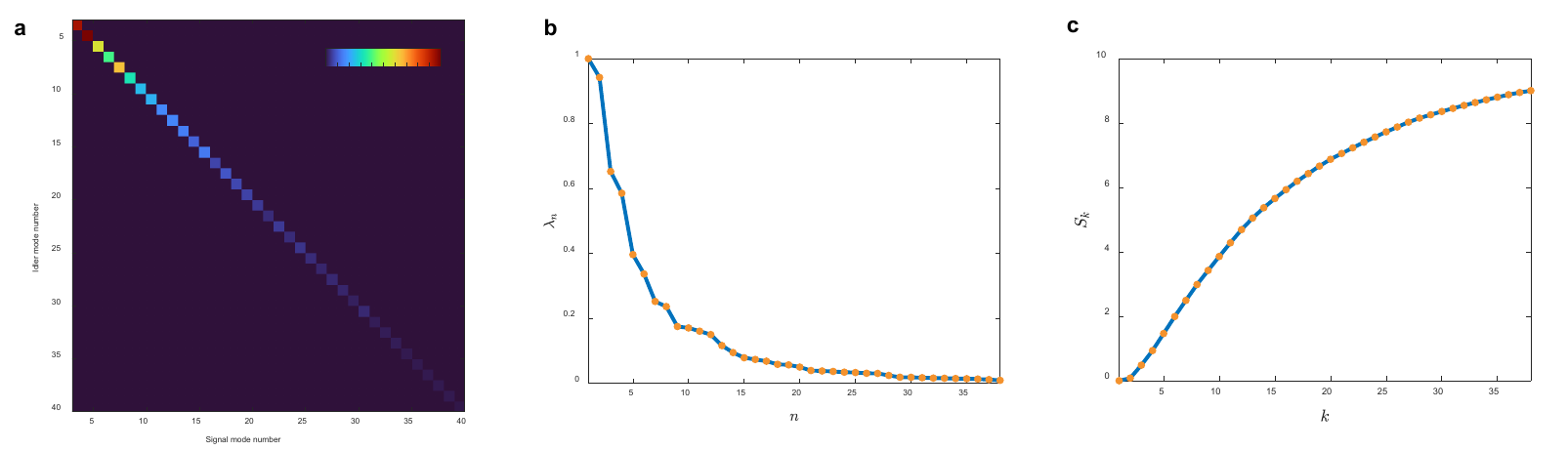}
\caption{\textbf{a} Simulated JSI of QFCs for the $\rm Si_{3}N_{4}$ resonator. \textbf{b} Schmidt coefficients $\lambda_n$ and \textbf{c} entropy of entanglement for the QFC.}
 \label{fig:S5}
\end{figure*}

This equation describes the frequency correlation (as a result of energy conservation $2\omega_{p}=\omega_{s}+\omega_{i}$) of signal and idler photons. Furthermore, the comb-like biphoton state can be considered a discretized result of continuous frequency entanglement~\cite{s9}. That is, the individual photon (a signal or idler photon) is a result of a superposition of hundreds of frequency modes, leading to a two-photon high-dimensional frequency-entangled state~\cite{s10}. Specifically, a signal (idler) photon generated from spontaneous FWM could be found in any signal (idler) frequency modes $|\omega_{p}-m\Delta\omega+\Omega\rangle$ ($|\omega_{p}+m\Delta\omega-\Omega\rangle$). And their emergence at corresponding frequency modes is highly correlated. These QFCs are also proved to be frequency-bin entangled~\cite{s11,s12} and energy-time entangled~\cite{s13,s14}.

By using Eq.~\ref{eq:s14},  we can assess the single-photon spectrum and JSI of generated QFCs. Generally, the single-photon spectrum of signal (idler) frequency modes can be given by~\cite{s8} 
\begin{equation}
S(\omega_{s})
=\langle\Psi|a_{s}^\dagger(\omega_{s})a_{s}(\omega_{s})|\Psi\rangle
=\eta^2\sum\limits_\mu\frac{\gamma_{s}\gamma_{i}\text{sinc}^2{(i\Delta k L)}}{|\gamma_{s}/2-i(\omega_{s}-\omega_{p}+m\Delta\omega)|^2|\gamma_{i}/2+i(\omega_{s}-\omega_{p}+m\Delta\omega)|^2},
\label{eq:s15}
\end{equation}
\begin{equation}
S(\omega_{i})
=\langle\Psi|a_{i}^\dagger(\omega_{i})a_{i}(\omega_{i})|\Psi\rangle
=\eta^2\sum\limits_\mu\frac{\gamma_{s}\gamma_{i}\text{sinc}^2{(i\Delta k L)}}{|\gamma_{s}/2-i(\omega_{p}-\omega_{i}+m\Delta\omega)|^2|\gamma_{i}/2+i(\omega_{p}-\omega_{i}+m\Delta\omega)|^2}.
\label{eq:s16}
\end{equation}

This equation reveals that bandwidths of the single-photon spectrum are highly related to the phase-matching condition. A carefully designed dispersion may lead to a spectral bandwidth of 250 THz, ranging from near-ultraviolet to mid-infrared~\cite{s15}. The generated frequency mode has a Lorentzian-like shape with a full width at half-maximum (FWHM) of $\sqrt{\sqrt{2}-1}\gamma\approx0.64\gamma$, where we consider $\gamma_\mathrm{s}=\gamma_1=\gamma$. For our resonator, we can calculate a peak FWHM of 74 MHz for signal (idler) frequency mode, where  cavity linewidth is $\gamma=115$ MHz.

Besides, we theoretically predict the JSI of generated QFCs by 
$\langle\Psi|a_{s}^\dagger(\omega_{s})a_{i}^\dagger(\omega_{i})a_{s}(\omega_{i})a_{i}(\omega_{i})|\Psi\rangle$.Therefore, the JSI can be given by
\begin{equation}
C_{JSI}(\mu) =\eta^2 \int_{-\infty}^{\infty}d\Omega\frac{\gamma_s\gamma_i}{(\gamma_s/2-i\Omega)^2(\gamma_i/2+i\Omega)^2}\text{sinc}^2{(\Delta k L)}
\label{eq:s17}
\end{equation}
Here, we can  obtain these parameters from the transmission spectrum shown in Fig.\ref{fig:S3}a, and ignore the term $2\Gamma P$. The normalized JSI of our $\rm Si_{3}N_{4}$ resonator is plotted in Fig.\ref{fig:S5}a.  Here we consider signal-idler mode numbers from 4 to 40 (modes 1–3 are significantly eliminated by the fiber Bragg grating (FBG) in the experiment).

To verify the frequency correlation of our biphoton frequency comb following the on-chip topological transport, we employ a coincidence count system to assess the JSI of QFCs. The entanglement exhibited by photon pairs can be elucidated by the factorizability of the JSA\cite{s9}. The JSA can be approximately obtained from the JSI by $\mathcal{A}(\omega_s,\omega_i)\approx\sqrt{|F(\mathcal{A}(\omega_s,\omega_i)|^{2}}$\cite{s10}.
We employ Schmidt decomposition to validate the entanglement present in the generated quantum frequency combs. If the biphoton state can be decomposed into a function of $\omega_s$ and $\omega_i$, it signifies the presence of high-dimensional frequency entanglement. The JSA can be expressed as follows\cite{s9}:

\begin{equation}
\mathcal{A}(\omega_{s},\omega_{i})=\sum_{n=1}^{N}\sqrt{\lambda_{n}}\psi_{n}(\omega_{s})\phi_{n}(\omega_{i}),
\label{eq:s18}
\end{equation}
where $\lambda_n$ ($N\in\mathbb{N}$) is denoted as Schmidt coefficient, $\psi_{n}$ and $\phi_{n}$ are are orthonormal functions of $\omega_s$ and $\omega_i$ in the Hilbert space. $\lambda_n$, $\psi_{n}$ and $\phi_{n}$ are connected by these equations

\begin{equation}
\begin{aligned}
\int K_1(\omega,\omega')\psi_n(\omega')d\omega'=\lambda_n\psi_n(\omega),\\
\int K_2(\omega,\omega')\phi_n(\omega')d\omega'=\lambda_n\phi_n(\omega),
\label{eq:s19}
\end{aligned}
\end{equation}
where $K_1$ and $K_2$ are so-called one-photon spectral correlations, and $\psi_{n}$ and $\phi_{n}$ are corresponding eigenfunctions. When the Schmidt number $N>1$, the biphoton state is frequency entangled. We can rewrite the equations as
\begin{equation}
\begin{aligned}
K_{1}(\omega,\omega^{\prime})=\int\mathcal{A}(\omega,\omega_{i})\mathcal{A}^{*}(\omega^{\prime},\omega_{i}){d}\omega_{i},\\
K_{2}(\omega,\omega^{\prime})=\int{\cal A}(\omega_{s},\omega){\cal A}^{*}(\omega_{s},\omega^{\prime}){d}\omega_{s},
\end{aligned}
\label{eq:s20}
\end{equation}
$K_{1}$ and $K_{2}$ form $s\times s$ and $i\times i$ matrices respectively.
The eigenfunctions can be represented as:
\begin{equation}
\begin{aligned}
K_1\psi_n=\lambda_n\psi_n,\\
K_{2}\phi_{n}=\lambda_{n}\phi_{n},
\end{aligned}
\label{eq:s21}
\end{equation}

Subsequently, Eq.\ref{eq:s18} can be restructured as:
\begin{equation}
{\cal A}=\sum_{n=1}^{N}\sqrt{\lambda_{n}}\psi_{n}\phi_{n}^{T},
\label{eq:s22}
\end{equation}

With Eq.\ref{eq:s22}, Schmit coefficients $\lambda_{n}$ can be computed by solving the eigenvalue equations. Notably, when the count of non-zero Schmidt coefficients $\lambda_{n}$ surpasses 1, or when the entanglement entropy $S_{k}>0$, the biphoton states exhibit frequency entanglement\cite{s9}. Moreover, the entropy of entanglement $S_k$ and Schmidt number $K$ serve as effective measures for quantifying the entanglement
\begin{equation}
S_k=-\sum\limits_{n=1}^N\lambda_n\log_2\lambda_n,
\label{eq:s23}
\end{equation}
\begin{equation}
K=-\frac{\left(\sum\limits_{n=1}^N\lambda_n\right)^2}{\sum\limits_{n=1}^N\lambda_n^2}.
\label{eq:s24}
\end{equation}
The entanglement of biphoton frequency combs is identified by $S_k>0$ or $K>0$, the large value of $S_k$ and $K$ leads to a high quality of high-dimensional frequency entanglement. For our $\rm Si_{3}N_{4}$ resonator, the theoretically calculated Schmidt number, entropy of entanglement are calculated as $K=11.40$ and $S_{k}=9$, respectively (Fig.\ref{fig:S5}b-c).
In the high-dimensional spaces, the Schmidt number $K$ is an effective metric for quantifying the degree of entanglement between signal and idler modes\cite{s10}. Therefore, the effective dimensions (numbers of relevant orthogonal modes) are larger than 11.
However, the Schmidt number calculated from experimental data is not large enough because of the performance limitations of InGaAs single photon detectors (SPDs).

\section{Numerical simulation of DKS combs}
\begin{figure*}[ht]
\centering
\includegraphics[width=0.9\textwidth]{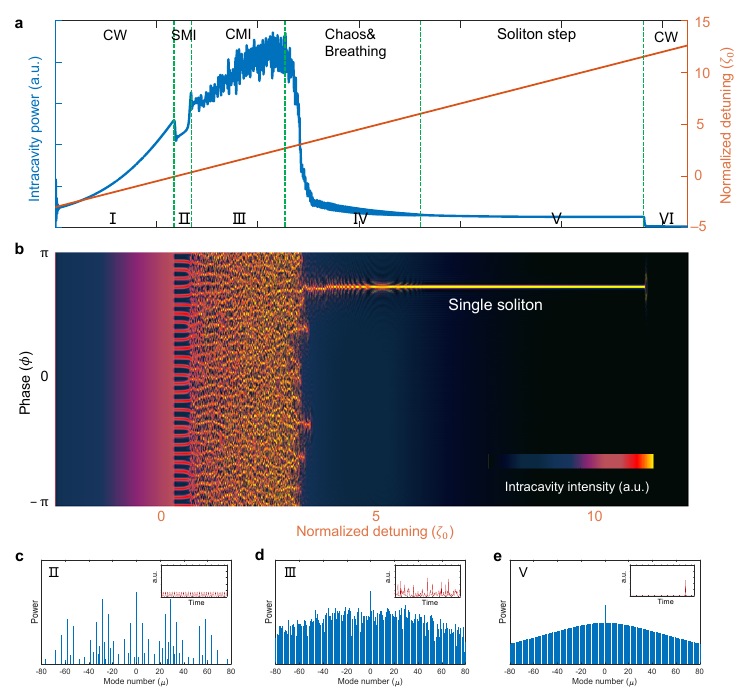}
\caption{A prototype for the generation of both QFCs and DKS combs.}
 \label{fig:S6}
\end{figure*}

The Lugiato–Lefever equation (LLE)~\cite{s16} describing the nonlinear evolution of the light field in micro-resonators can be given by the nonlinear Schrödinger equation (NSE):
\begin{equation}
\frac{\partial}{\partial z}A^{(m)}+\frac{\alpha}{2}A^{(m)}+i\frac{\beta_{2}}{2}\frac{\partial^{2}}{\partial T^{2}}A^{(m)}=i\gamma\big|A^{(m)}\big|^{2}A^{(m)},
\label{eq:S25}
\end{equation}	
where $A^{(m)}$ is the field envelope for $m$-th roundtrips, $L$ is length of the micro-resonator. The boundary condition can be written as 
\begin{equation}
A^{(m+1)}(0,T)=\sqrt{\Theta}A_i+\sqrt{1-\Theta}\exp(-i\delta_0)A^{(m)}(L,T),
\label{eq:S26}
\end{equation}	
where $T$ represents the fast time variable that describes the waveform, $A_i$ is the pump field, $\Theta$ and $\delta_0$ are the coupling coefficient and detuning of the resonance frequency, respectively. $\alpha$, $\beta_{2}$, and $\gamma$ are the roundtrip loss, second-order dispersion term, and nonlinear coefficient, respectively. 
Assuming that the light field changes very little after a propagating distance of $L$, then ${\partial}/{\partial z}$ can be replaced by a slope with
\begin{equation}
\frac{\partial}{\partial z}A^{(m)}(z,T)\bigg|_{z=0}=\frac{A^{(m)}(L,T)-A^{(m)}(0,T)}{L}.
\label{eq:S27}
\end{equation}	
If the light field does not travel through the coupling region,  substituting the Eq.~\ref{eq:S27} into Eq.~\ref{eq:S25}, then we have 
\begin{equation}
A^{(m)}(L,T)=A^{(m)}(0,T)+L\left(-\frac{\alpha}{2}A^{(m)}(0,T)-\frac{i\beta_{2}}{2}\frac{\partial^{2}}{\partial T^{2}}A^{(m)}(0,T)+i\gamma\Big|A^{(m)}(0,T)\Big|^{2}A^{(m)}(0,T)\Big).\right. 
\label{eq:S28}
\end{equation}	
The NSE gives how the light field changes when it travels a distance of $L$.
Then we  assume the power coupling coefficient is far smaller than 1, that is, $\Theta\ll 1$, and the detuning is  far smaller than FSR, that is, $\delta_{0}\ll2\pi$. We can rewrite the Eq.~\ref{eq:S26} as  
\begin{equation}
A^{(m+1)}(0,T)=\sqrt{\Theta}A_i+\left(1-\frac{\Theta}{2}-i\delta_0\right)A^{(m)}(L,T).
\label{eq:S29}
\end{equation}	
To reduce the complexity of derivations, we replace the term m with slow time variable $t_R$, where increasing time $t_R$ (roundtrip time) denotes the increasing of roundtrips m. Therefore, we can obtain the relation
\begin{equation}
\frac{\partial}{\partial\tau}A(\tau,T)=\frac{A^{(m+1)}(0,T)-A^{(m)}(0,T)}{t_{R}}.
\label{eq:S30}
\end{equation}	
We can rewrite the Eq.~\ref{eq:S30} with new symbolic expression:
\begin{equation}
t_R\frac{\partial}{\partial\tau}A=-\left(\frac{\alpha L+\Theta}{2}+i\delta_0\right)A-iL\frac{\beta_2}{2}\frac{\partial^2}{\partial T^2}A+iL\gamma|A|^2A+\sqrt{\Theta}A_i.
\label{eq:S31}
\end{equation}	
This is the first form of LLE. Consequently, we replace several mathematical expressions to make the LLE more understandable. For example, the roundtrip time $t_R$ can be calculated from the FSR of the resonator by $t_R=1⁄FSR$, the intrinsic loss and external loss can be expressed as $\kappa_{in}=\alpha L\cdot FSR$ and $\kappa_{ex}=\Theta\cdot FSR$, respectively. The total loss $\kappa=\kappa_{in}+\kappa_{ex}$ corresponds to the resonance linewidth.  The normalized detuning is
\begin{equation}
\delta_{0}=\beta_{1}L\big(\omega_{0}-\omega_{p}\big)=\frac{1}{FSR}\delta\omega,
\label{eq:S32}
\end{equation}	
where $\delta\omega$ is the detuning $\omega_{0}-\omega_{p}$. The second-order dispersion is given by
\begin{equation}
D_{2}=-\frac{L}{2\pi}\beta_{2}(2\pi f_{r})^{3}.
\label{eq:S33}
\end{equation}	
The fast time variable $T$ is replaced by the microcavity angular coordinate $\phi$:
\begin{equation}
T=\frac{1}{2\pi f_{r}}\phi.
\label{eq:S34}
\end{equation}	
Therefore, we can get the second form of LLE:
\begin{equation}
\begin{aligned}\frac{\partial}{\partial\tau}A&=-\left(\frac{\kappa}{2}+{i}\delta\omega\right)A+{i}\frac{D_{2}}{2}\frac{\partial^{2}}{\partial\phi^{2}}A+{i}Lf_{r}\gamma|A|^{2}A+\sqrt{f_{r}\kappa_{{ex}}}A_{i}.\end{aligned}
\label{eq:S35}
\end{equation}	
where $\tau$ is the slow time variable, $f_{r}$ is the FSR of the resonator. By using the  Eq.~\ref{eq:S35}, we can numerically simulate the nonlinear dynamic evolution of the Kerr solitons in our $\rm Si_{3}N_{4}$ resonator. 
In our simulation, the pump power is set as 0.4 W, the FSR of the resonator is 95.75 GHz,  the nonlinear index is $2.5\times10^{-19}$ $\rm {m}^{2}{W}^{-1}$, the second-order dispersion is $D_2=5.95\times10^6$ rad/s, and the $Q$-factor is $1.68\times10^{6}$. 
We set a simulated effective field cross-section area at the pump wavelength by $A_{eff}=2.1\times10^{-14}$ $\rm{m}^{2}$, and therefore the nonlinear coefficient can be given by $\gamma=\omega_{0}n_{2}/cA_{eff}$.

The simulated intracavity energy and the corresponding spatiotemporal evolution of DKS combs as a function of the detuning are depicted in Fig.\ref{fig:S6}a-b.
We can clearly see several states, including stable modulation instability (SMI), chaotic modulation instability (CMI), breathing, and soliton states. The solitons always exist at the red-detuned side of the resonance frequency, where the intracavity field is bistable. 
The simulated optical frequency combs are shown in Fig.\ref{fig:S6}c-e, the result reveals the existence of a single soliton state.

\section{RF beatnotes of the single soliton comb}
\begin{figure*}[h]
\centering
\includegraphics[width=1\textwidth]{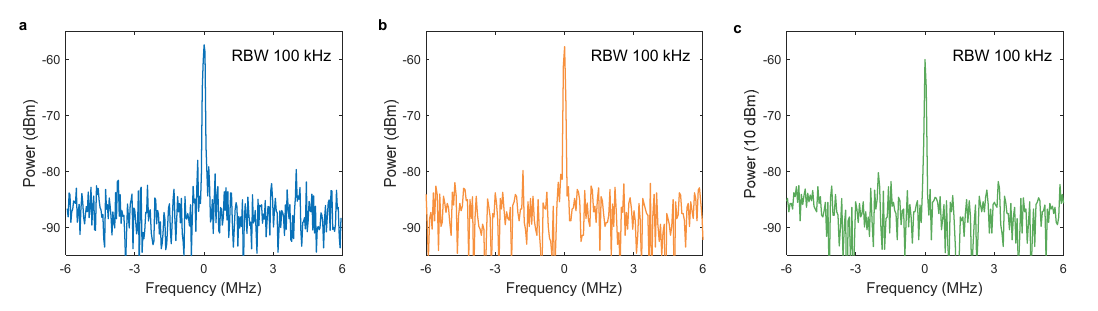}
\caption{RF beatnotes of the single soliton states for the \textbf{a} original DKS combs, DKS combs after the transport of the \textbf{b} straight and \textbf{c} Z-shaped topological waveguides, respectively.}
 \label{fig:S7}
\end{figure*}

To further characterize the performance of the single soliton comb, a reference CW laser (TSL) is employed to generate a single-wavelength laser with a typical linewidth of 60 kHz. The output combs are heterodyned with a CW laser and then directed to a photodetector. The resulting electrical spectrum was measured with an electrical spectrum analyzer, as shown in Fig.~\ref{fig:S7}. The resolution bandwidths (RBWs) are around 100 kHz. The signal-to-noise ratio was approximately 30 dBm, indicating the presence of a narrow pulse width in this configuration of the single soliton comb.

%